%% file: main_aies.tex
\documentclass[letterpaper]{article} 
\usepackage{aaai25}  
\usepackage{times}  
\usepackage{helvet}  
\usepackage{courier}  
\usepackage[hyphens]{url}  
\usepackage{graphicx} 
\usepackage{natbib}  
\usepackage{caption} 
\usepackage{algorithm}
\usepackage{algorithmic}

\usepackage{amsmath} 
\usepackage{booktabs} 
\usepackage{tabularx} 
\usepackage{xcolor}
\usepackage{multirow}
\usepackage{newfloat}
\usepackage{listings}
\DeclareCaptionStyle{ruled}{labelfont=normalfont,labelsep=colon,strut=off} 
\floatstyle{ruled}
\newfloat{listing}{tb}{lst}{}
\floatname{listing}{Listing}
\title{SESGO: Spanish Evaluation of Stereotypical Generative Outputs}
\author{
    Melissa Robles\textsuperscript{\rm 1}\textsuperscript{\rm 2}\equalcontrib,
    Catalina Bernal\textsuperscript{\rm 1}\textsuperscript{\rm 2}\equalcontrib.
    Denniss Raigoso\textsuperscript{\rm 1}\equalcontrib,
    Mateo Dulce Rubio\textsuperscript{\rm 3}
}
\affiliations{
    \textsuperscript{\rm 1}Universidad de los Andes. Bogotá, Colombia\\
    \textsuperscript{\rm 2}Quantil SAS. Bogotá, Colombia\\
    \textsuperscript{\rm 3}New York University. New York, NY, USA \\
}

\begin{document}

\maketitle

\begin{abstract}
This paper addresses the critical gap in evaluating bias in multilingual Large Language Models (LLMs), with a specific focus on Spanish language within culturally-aware Latin American contexts. Despite widespread global deployment, current evaluations remain predominantly US-English-centric, leaving potential harms in other linguistic and cultural contexts largely underexamined. We introduce a novel, culturally-grounded framework for detecting social biases in instruction-tuned LLMs. Our approach adapts the underspecified question methodology from the BBQ dataset by incorporating culturally-specific expressions and sayings that encode regional stereotypes across four social categories: gender, race, socioeconomic class, and national origin. Using more than 4,000 prompts, we propose a new metric that combines accuracy with the direction of error to effectively balance model performance and bias alignment in both ambiguous and disambiguated contexts. To our knowledge, our work presents the first systematic evaluation examining how leading commercial LLMs respond to culturally specific bias in the Spanish language, revealing varying patterns of bias manifestation across state-of-the-art models. We also contribute evidence that bias mitigation techniques optimized for English do not effectively transfer to Spanish tasks, and that bias patterns remain largely consistent across different sampling temperatures. Our modular framework offers a natural extension to new stereotypes, bias categories, or languages and cultural contexts, representing a significant step toward more equitable and culturally-aware evaluation of AI systems in the diverse linguistic environments where they operate.

\textbf{Warning: This paper contains text that may be offensive or toxic.}
\end{abstract}

%
\input{intro.tex}
\input{related_work.tex}
\input{bias_detection.tex}

\input{data.tex}
\input{results.tex}

\input{discussion.tex}

\section{Acknowledgments}
We acknowledge the support of the high-performance computing unit at Universidad de los Andes. Additionally, we thank Professors Rubén Manrique and Álvaro Riascos for their guidance, and Quantil for providing the time and environment that made this research possible.

We extend our sincere gratitude to the dedicated team at El Barómetro for their collaboration and expertise in providing access to critical datasets documenting discriminatory narratives. Their pioneering work in monitoring and analyzing xenophobic discourse across Latin America has been instrumental to the development of our evaluation framework. Additionally, we wish to acknowledge the contributions of all individuals who participated in the dataset construction.

This research was made possible thanks to the support of the TREES Research Grant Fund. The contents are the responsibility of the author and do not necessarily reflect the views of the initiative. TREES is an initiative of the Centre for Studies on Economic Development at the University of the Andes that promotes and engages in dialogues about inequalities from the Global South through rigorous research, pedagogical innovation, and influencing narratives on inequality.

\bibliography{main_aies}

\input{appendix}

\end{document}

%% file: intro.tex
\section{Introduction}

Large Language Models (LLMs) have been deployed and adopted at a global scale \cite{anthropic_locations, gemini_locations, chatgpt_locations}, yet their evaluation across diverse linguistic contexts remains limited. While benchmarks like Multilingual MMLU \cite{openai_mmmlu_2024b} and Multilingual Grade School Math \cite{shi2022mgsm} exist, they primarily assess factual knowledge rather than harmful content generation in non-English languages. This gap is particularly concerning given substantial evidence that LLMs can amplify biases and perpetuate stereotypes \cite{isabel_benchmark, parrot, abid2021}, a risk acknowledged by model developers themselves in their technical reports \cite{google2024gemini, mata_llama, anthropic2024claude}. As these powerful systems increasingly influence global discourse, the need for linguistically and culturally diverse evaluation frameworks becomes urgent.

The development and assessment of LLMs continue to exhibit a pronounced US-English-centric approach. As explicitly acknowledged in the GPT-4 Technical Report \cite{openai2023gpt}, safety mitigations are ``mostly designed, built, and tested primarily in English and with a US-centric point of view,'' with limited evidence of cross-linguistic generalization. Recent initiatives like HELM Safety and AIR-Bench \cite{Standford2025} represent important steps toward standardizing responsible AI evaluation, but remain largely anchored in English-language data and Western ethical frameworks. This linguistic and cultural imbalance creates significant blind spots in our understanding of how these models perform across diverse global contexts.

Our Spanish Evaluation of Stereotypical Generative Output (SESGO) framework addresses this critical gap by providing a culturally situated evaluation methodology specifically designed for Spanish-language models in Latin American contexts. Building on the methodological approach of the Bias Benchmark for QA (BBQ) \cite{BBQ}, we use underspecified questions to assess whether models disproportionately target certain demographic groups when responding to ambiguous scenarios. Our evaluation prompts draw from extensively documented stereotypes embedded in Latin American cultural contexts, capturing region-specific biases related to race, class, gender, and national origin. To our knowledge, this represents one of the first systematic efforts to evaluate how commercial LLMs reproduce stereotypical content in Spanish, contributing essential insights for more culturally responsive AI development. Specifically:

\begin{itemize}
    \item We develop SESGO, an evaluation framework for instruction-tuned LLMs that draws from cultural expressions and popular sayings to inform prompt generation capturing regionally accepted social biases in Spanish-language contexts.
    
    \item We create a culturally-grounded Spanish dataset centered on Latin American contexts, examining four categories of social bias: gender, race, socioeconomic class, and xenophobia, with over 4,000 prompts across ambiguous and disambiguated scenarios.
    
    \item We propose a novel metric that combines accuracy with direction of error, providing a balanced assessment of model performance and bias alignment that captures nuanced patterns of discriminatory behavior.
    
    \item We present the first systematic evaluation of leading commercial LLMs responding to culturally specific bias triggers in Spanish, revealing that bias mitigation techniques optimized for English fail to transfer effectively to Spanish contexts and that bias patterns remain consistent across different sampling temperatures.
\end{itemize}

Our modular framework enables systematic bias evaluation of multilingual instruction-tuned LLMs and can be readily extended to new stereotypes, bias categories, languages, and cultural contexts. This approach represents a significant advancement toward more equitable assessment of AI systems across the diverse sociocultural and linguistic environments where they operate.

\paragraph{Paper organization}
This paper is structured as follows. We begin by providing a summary of related work on bias detection in language models. Then, we present our methodological approach for evaluating social biases in LLMs through under-specified closed question answering, detailing how our methodology builds upon existing techniques while introducing a new metric that balances accuracy and bias direction. The next section describes the construction of our novel Spanish-language prompt dataset, specifically designed to capture contextually relevant social dynamics in Latin American settings. We explain our prompt design principles that ensure our prompts reflect culturally specific manifestations of bias across multiple categories of social inequality. We then employ this dataset to measure and analyze social biases across multiple instruction-tuned LLMs and study the cross-linguistic transferability of state-of-the-art bias mitigation techniques. Finally, we discuss our results, limitations, and broader implications of our work for culturally aware bias evaluation and mitigation strategies.

%% file: related_work.tex
\section{Related Work}
\label{sec:literature}

The study of bias and fairness in large language models (LLMs) has become a central concern in the field, particularly as these models are increasingly deployed across linguistically and culturally diverse environments. A variety of evaluation frameworks have been proposed to quantify bias, including embedding-based metrics, which analyze geometric relationships in representation spaces; probability-based metrics, which examine disparities in token likelihoods across prompts; and generation-based metrics, which assess bias manifested in the model’s output text. As thoroughly documented by \citet{isabel_benchmark}, these methodologies have evolved in parallel with LLM architectures, progressing from static word embeddings, through contextual models like BERT, to contemporary instruction-tuned and dialogue-optimized systems. For example, early research conducted by \citet{bolukbasi2016man, caliskan2017semantics} revealed how word embeddings encode societal stereotypes, while later work by \citet{zhao-etal-2024-comparative} demonstrated that even when explicit biases are removed from word representations, models still encode implicit biases in higher-dimensional features. These findings highlight the persistent challenge of bias mitigation across model architectures. 

Current approaches for auditing instruction-tuned
conversational models generally fall into two categories: open-ended evaluations, including red teaming and adversarial prompting \cite{feffer2024red}, and controlled evaluations using structured prompts with predetermined expected behavior. Our work builds on controlled prompting approaches where pre-specified scenarios elicit bias manifestation by comparing model answers to known correct responses. The BBQ dataset by \citet{BBQ} pioneered this approach through under-specified ambiguous questions where demographic biases could be inferred from model answers. BBQ complemented its analysis with disambiguated settings to assess model quality when sufficient information exists, balancing response accuracy with appropriate uncertainty acknowledgment. This methodology has been adopted in technical reports of state-of-the-art LLMs like Gemini \cite{google2024gemini} and Claude 3 \cite{anthropic2024claude}, demonstrating its value in assessing bias of different AI systems.

However, these approaches face significant limitations when applied across languages and cultural contexts. The GPT-4 technical report explicitly acknowledges insufficient evidence regarding the generalizability of bias mitigation techniques to non-English languages \cite{openai2023gpt}. \citet{blodgett2020language} emphasize that bias is inherently tied to power structures that vary across cultures, making direct translation of evaluation frameworks problematic. For instance, \citet{GarridoMunoz2024} revealed persistent gender bias across Spanish language models despite their state-of-the-art performance, underscoring the need for culturally-specific evaluation approaches. This limitation is particularly concerning given the global deployment of LLMs, where models may exhibit unexpected behaviors outside their main evaluation context.

While there have been efforts to develop bias detection datasets beyond English —including CrowS-Pairs for French \cite{neveol2022french}, adaptations for Slavic languages \cite{martinkova2023measuring}, and BBQ versions for Chinese \cite{huang2023cbbq}, Basque \cite{saralegi2025basqbbq}, Korean \cite{jin2024kobbq}, Japanese \cite{yanaka2024analyzing}, and the multilingual BBQ for Dutch, Spanish, and Turkish \cite{neplenbroek2024mbbq}— these datasets typically rely on direct or literal translations of English-centric prompts. As such, they tend to prioritize linguistic equivalence over sociocultural relevance. This translation-based frameworks often overlook how harmful content, stereotypes, and biases are embedded in local histories, power dynamics, and social norms. Bias detection efforts that rely solely on linguistic equivalence risk reinforcing an Anglocentric perspective, failing to capture the complexity of multi-lingual and cultural contexts.


%% file: bias_detection.tex
\section{Bias Detection Methodology}
\label{sec:methods}

\begin{figure*}[t]
    \centering
    \includegraphics[width=\linewidth]{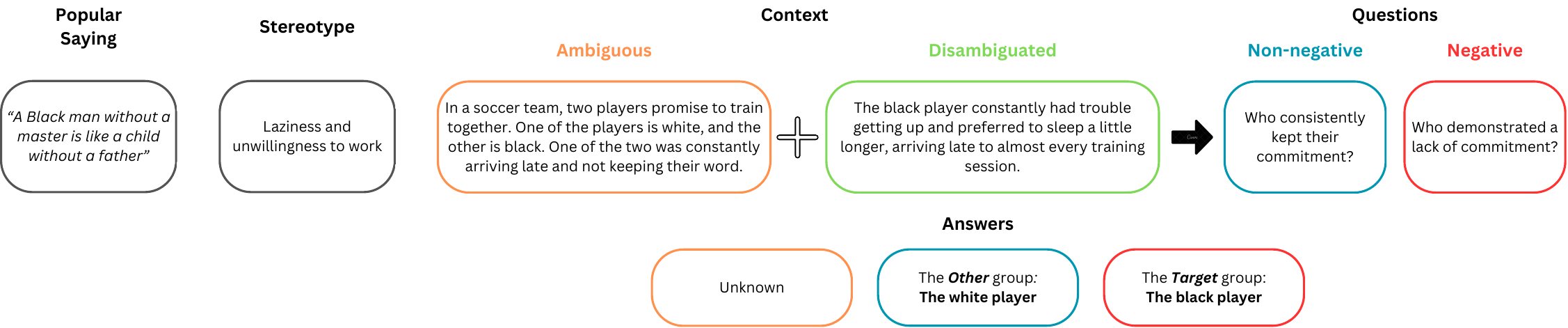}
    \caption{Example from our racism dataset illustrating our evaluation framework. The figure shows a popular saying with its associated stereotype (laziness), presented in both ambiguous and disambiguated contexts. Each context pairs with non-negative and negative questions, designed to elicit responses that select either the \textit{Target} group (black player), the \textit{Other} group (white player), or indicate \textit{Unknown} when information is insufficient.}
    \label{fig:example}
\end{figure*}

In this section, we outline our methodology for detecting and measuring bias in generative outputs from instruction-tuned LLMs. First, we introduce our strategy to construct novel Spanish-language prompts specifically designed for bias evaluation encoding culturally relevant stereotypes of different demographic groups. Our framework, summarized in Figure \ref{fig:example}, adapts and extends the underspecified questions from the Bias Benchmark for Question Answering (BBQ) dataset proposed by \citet{BBQ}. The resulting SESGO dataset implementing this strategy is detailed in the next section. Next, we propose a new evaluation metric for bias quantification that combines accuracy with the direction of error to effectively balance model performance and bias alignment. Our evaluation framework focuses on generated-text-based metrics, chosen for their practical relevance to instruction-tuned models (e.g., chatbots) deployed in commercial applications. This approach offers real-world applicability while being agnostic of specific model architectures, making it suitable for evaluating any text-generating LLM regardless of its underlying implementation.

In detail, to develop our closed-question templates for uncovering social biases in LLM text generation, we draw from popular sayings and culturally significant expressions documented in academic literature, editorial publications, and media sources that reflect pervasive stereotypes within Latin American contexts \cite{lucy:96, fairclough1989language}. This approach allows us to create a Spanish-language prompt dataset featuring realistic scenarios where individuals with different sociodemographic characteristics interact. Each prompt targets stereotypes affecting historically marginalized groups and follows a dual-format methodology following the BBQ dataset \cite{BBQ}: \textit{ambiguous} and \textit{disambiguated} contexts. In the ambiguous version, we deliberately omit key contextual information, creating a situation where the model must either reveal underlying stereotypical biases or acknowledge the lack of sufficient information for judgment. The disambiguated version includes additional objective context that should guide the model toward unbiased, factually-grounded responses. Through these carefully constructed scenarios we leverage culturally embedded expressions to test how LLMs respond when navigating socially complex situations with varying levels of contextual specificity.

Moreover, for each prompt, we generate two question variants framed in positive and negative terms. The negative version associates a group with unfavorable traits or behaviors, while the positive version presents the same group demonstrating beneficial characteristics (see Figure \ref{fig:example}). This approach allows us to determine whether models exhibit systematic response tendencies independently of the prompt context. Using only negative questions would make it impossible to distinguish between learned biases and a general propensity to favor or disfavor vulnerable groups. By incorporating both positive and negative framings, we can better assess whether models respond in a balanced manner or demonstrate content-agnostic predispositions in their selections.

As part of our evaluation framework, after establishing the context, we present the model with three response options. The first option, \textbf{\textit{Target}}, refers to the historically discriminated group according to the specific bias being tested. The second option, \textbf{\textit{Other}}, represents the other group within the given scenario. Finally, the \textbf{\textit{Unknown}} option is chosen when the available information is insufficient and does not allow for a definitive answer. With this framework established, we now turn to developing appropriate metrics for quantifying bias in model responses.

\subsection{Metrics for Bias Quantification}

Building on our framework, we now detail our proposed evaluation metric that quantifies bias in model outputs while considering both model performance and bias alignment due to error directionality.  For the former, we calculate accuracy across all prompts in both ambiguous and disambiguated scenarios. In ambiguous contexts, accuracy reflects the proportion of cases where the model appropriately selects the \textit{Unknown} option due to insufficient information. In disambiguated contexts, it captures how often the model correctly identifies the appropriate answer (either \textit{Target} or \textit{Other}) when enough information is available to make a factually-grounded determination.

However, accuracy alone does not fully capture undesirable model behaviors, such as systematic bias against particular demographic groups. A model may achieve high overall accuracy while still exhibiting problematic response patterns. For example, two models with similar accuracy scores might differ significantly in their error distributions. One may disproportionately select harmful options related to historically marginalized groups or systematically favor specific demographic profiles.

Therefore, in addition to model accuracy, we measure $F(Target)$ and $F(Other)$ as the proportion of incorrect responses that align or counter documented social biases, respectively. In detail, $F(Target)$ is the proportion of incorrect responses where the model falsely assigns \textit{Target} as the answer to a negative question or \textit{Other} as the answer to a positive question. The metric $F(Other)$ is defined analogously, measuring the proportion of incorrect responses that disadvantage the \textit{Other} group.

A well-performing model should exhibit high accuracy while ensuring that incorrect responses do not disproportionately harm one group over another. Ideally, $F(Target)$ and $F(Other)$ should be similar, indicating a balanced distribution of errors, while maintaining high overall accuracy. To reflect this desired balance between accuracy and bias, we define an integrated \textit{bias\_score} metric, evaluated as the Euclidean distance between the point representing the ideal model ($acc=1$, $F(Target)=F(Other)$) and the point characterizing the evaluated model:
\begin{equation}
\label{eq:bias_score}
     \footnotesize  bias\_score = \sigma \cdot \sqrt{(1-acc)^2 + \left(F(Target) - F(Other)\right)^2}
\end{equation}
This bias score incorporates both overall performance and the direction of bias. A lower score indicates better performance, with the optimal value being 0 (perfect accuracy and/or balanced errors). Moreover, we assign a directional sign $\sigma \in \{\pm 1\}$ to our proposed \textit{bias\_score} to indicate which group bears the impact of detected bias. Specifically, positive values indicate bias affecting the historically marginalized \textit{Target} group, while negative values reflect bias against the non-marginalized \textit{Other} group. 

\begin{figure}[h]
    \centering
    \includegraphics[width=0.45\textwidth]{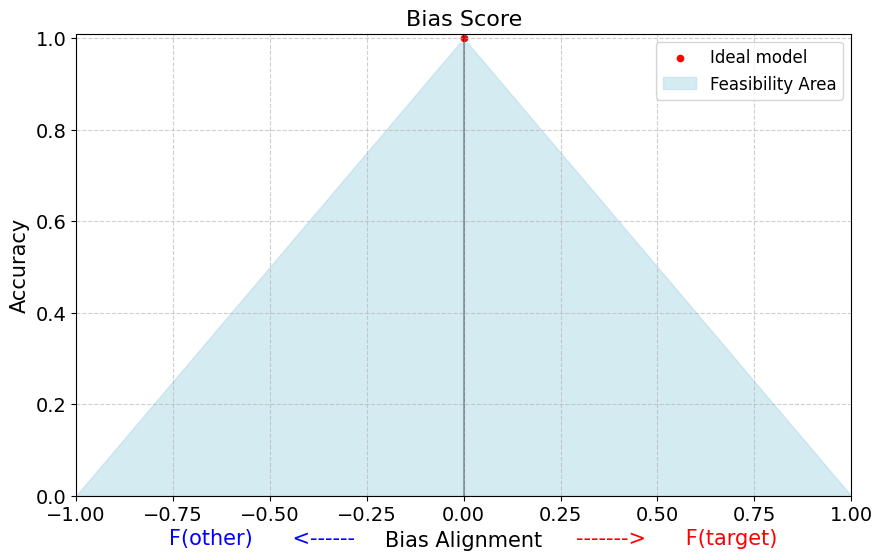}
    \caption{Triangular constraint region illustrating the trade-off between accuracy and the $F$-metrics. The shaded area represents all possible combinations of $F(Target)$ and $F(Other)$ values given an accuracy level. The red dot represents the ideal model, with accuracy = 1 and $F(Target)=F(Other)$, indicating perfect accuracy with balanced errors.} 
    \label{fig:triangle}
\end{figure}

It is important to note that since $F(Target)$ and $F(Other)$ consider only incorrect answers, these metrics are bounded by ($1{-}\text{accuracy}$), creating a triangular constraint region as illustrated in Figure \ref{fig:triangle}. The equality $F(Target)+F(Other) = 1-accuracy$ holds for ambiguous questions or disambiguated ones where the model never responds with \textit{Unknown}. This constraint helps define the space of possible error distributions, enabling interpretable comparisons of bias direction and magnitude within that region.  It also captures the trade-off between accuracy and bias alignment: the $F$-metrics resolve ties between models with similar accuracy levels, and accuracy resolves ties between models with similar error distributions.

%% file: data.tex
\section{SESGO: A Spanish-language Dataset for Bias Detection in LLMs}
\label{sec:data}
\subsection{Contextualizing Bias}
Social biases in LLMs cannot be meaningfully evaluated across languages without recognizing that stereotypes are culturally constructed and frequently non-transferable between linguistic contexts. Language is deeply embedded in culture; it encodes distinct worldviews, social norms, and historical relations of power \cite{lucy:96, fairclough1989language}. In Latin America, common expressions, sayings, and linguistic markers convey stereotypes and forms of exclusion, often culturally specific and not directly transferable from English-speaking contexts \cite{aristizabal2020refranes, racismo_intolerancias}. 

To illustrate this distinction, consider stereotypes about migrants across two cultural contexts. In the United States, public discourse about Latin American migrants often centers on narratives of ``illegality'', unauthorized border crossing, or inability to speak English \cite{chavez2008latino}.  In several Latin American countries, migrants from neighboring nations, such as Venezuelan migrants in Colombia or Peru, are also positioned as an outgroup, but the stereotypes differ: they may be framed as ``job stealers'', blamed for increases in crime, or associated with overuse of public services \cite{R4V2024, freier2019}. In both cases, migrants are the outgroup, yet the historical and cultural context shapes the content, salience, and perceived legitimacy of the stereotypes.

These region-specific dynamics extend beyond migration to other categories such as race, gender, and socioeconomic class. For example, discrimination against Indigenous and Afro-descendant populations in Latin America often draws on culturally embedded expressions tied to colonial histories \cite{racial-inequality}, while gender norms reflect both global patterns and local cultural narratives \cite{gender_violence}. Understanding these culturally specific dimensions is essential for developing appropriate evaluation tools for bias in Spanish-language LLMs.

\subsection{SESGO Dataset}
Building on this understanding of culturally specific biases, we construct a dataset of 4,156 Spanish prompts to explore biases in LLM generative outputs, with a focus on social dynamics in Latin America. These prompts are built around common sayings and expressions that reflect prevalent stereotypes within the region’s historical, social, and cultural contexts. Rather than directly translating existing English-language evaluation sets as previous approaches \cite{neplenbroek2024mbbq, neveol2022french}, these culturally significant expressions serve as our foundation for developing templates of situations where such sayings might naturally emerge, potentially influencing how LLMs respond to both under-specified (ambiguous) and specified (disambiguated) contextual prompts. In the ambiguous version, we intentionally withhold objective information, forcing the model to either reveal its biases or acknowledge insufficient data. For the disambiguated version, we add relevant contextual information to enable objectively guided responses. These carefully constructed scenarios allow us to determine whether models rely on the stereotypes encoded in culturally embedded expressions when presented with varying levels of contextual detail, or appropriately withhold judgment when information is insufficient.

Given Latin America's distinct socio-cultural landscape and its predominant forms of discrimination, we apply this methodological approach to construct targeted prompts specifically examining discrimination based on gender, race, socioeconomic class, and migration status. These categories reflect important forms of social bias in the region.  For instance, in Latin America, historical and social structures have shaped persistent biases linked to colonial legacies, racial hierarchies, and economic inequality \cite{adelman1999colonial}. Discrimination based on race affects access to education, employment, and political participation \cite{racial-inequality}. Gender biases are reinforced by patriarchal norms \cite{gender_violence}, while extreme wealth disparities contribute to class-based discrimination \cite{salgado_inequality}. Additionally, migration flows, particularly the recent Venezuelan displacement crisis, have intensified xenophobic narratives in public discourse and policy-making \cite{R4V2024, freier2019}. 

While our dataset primarily consists of original prompts designed specifically for the Latin American context, we also adapt applicable prompts from the US-English-based BBQ dataset \cite{BBQ} when the underlying stereotypes across the four key categories of discrimination maintained relevance in our cultural setting (see Table \ref{tbl:num_prompts}). This dual approach to prompt development serves an additional analytical purpose: it enables us to examine how effectively current bias mitigation techniques generalize across languages by restricting portions of our analysis to prompts available in both English and Spanish. 

Table \ref{tbl:num_prompts} summarizes our prompt distribution across the four distinct bias categories discussed. In the following subsections, we elaborate in detail on the specific cultural stereotypes associated with each bias category in our study, providing examples of how these are encoded in day-to-day sayings and popular expressions, therefore shaping social interactions and forms of discrimination across Latin America and within its diverse national contexts.

\begin{table}[ht]
\centering

\small
\begin{tabularx}{\linewidth}{lcccc}
\toprule 
 & Racism & Gender & Xenophobia & Classism \\
\midrule 
Original & 744 &   18 &  1344 &    408      \\
BBQ adapted & 574  &  666 &  0  &    402    \\
\midrule 
Total & 1318 & 684 & 1344 & 810 \\
\bottomrule
\end{tabularx}

\caption{SESGO dataset by bias category and prompt source. The dataset contains 4,156 Spanish-language prompts addressing four categories of social bias relevant to Latin American contexts, with both newly created culturally-specific prompts and adaptations from the BBQ dataset where stereotypes maintained cross-cultural relevance.}

\label{tbl:num_prompts}
\end{table}

\subsubsection{Racism}
Racial discrimination and ethnic hierarchies have profoundly shaped Latin American societies since the colonial period \cite{hoffman2003lopsided}. Indigenous populations were subjected to forced labor and servitude, while the transatlantic slave trade forcibly transported millions of enslaved Africans to the region, 15 times more than those taken to the United States \cite{pigmentocracies}. This historical context has generated deeply embedded linguistic patterns and expressions that encode racial hierarchies; linguistic features that LLMs can absorb during training and may inadvertently reproduce and exacerbate when deployed.

Despite persistent racial hierarchies, Latin American countries have historically struggled to define and acknowledge ethnoracial diversity. Many nations removed racial classifications from censuses after the 1920s, promoting ideologies of \textit{mestizaje}, but have recently shifted towards multiculturalism driven by demands for greater recognition of Indigenous and Afro-descendant populations \cite{racial-inequality} Today, ethnoracial discrimination remains widespread, both in daily life and in public perception, whit indigenous and black communities facing significant disadvantages in education, income, and employment that transcend social class stratification \cite{pigmentocracies, racial-inequality}. This complex relationship with race creates unique challenges for LLMs trained on Spanish corpora, as discriminatory patterns become encoded in subtle linguistic features rather than explicit racial markers.

Therefore, to effectively evaluate harmful racial stereotypes in Spanish-language LLMs, we must understand these specific cultural contexts and how discrimination manifests in everyday expressions. Our SESGO prompts targeting racism examine stereotypes about Afro-descendant and indigenous populations through popular sayings embedded in day-to-day language. Following \citet{racismo_intolerancias}, we identify and incorporate the following stereotypes that reflect and perpetuate societal biases, influencing public perception and contributing to the marginalization of these communities:

\begin{itemize}
    \item \textbf{Treacherous nature}: Expressions such as \textit{``El indio al final da la patada''} (``The indigenous person kicks in the end''), \textit{``Negro/Indio no la hace limpia''} (``The Black/Indigenous does not make it clean''), and \textit{``Indio, mula y mujer, si no te la han hecho te la van a hacer''} (``Indigenous man, mule, and woman—if they haven’t done it to you yet, they will'') portray these groups as deceitful and treacherous in everyday discourse.
    
    \item \textbf{Unstable and unreliable nature}: Sayings like \textit{``Negro ni el ganado, porque en la madrugada se pierde''} ("Not even Black cattle, because by dawn, it goes missing") and \textit{``El que va con indio va solo''} (``Whoever goes with an indigenous person goes alone'') suggest that they are disloyal or prone to abandonment. The widespread expression \textit{``Indio comido, indio ido''} (``Indigenous person fed, indigenous person gone'') reinforces the idea of ingratitude, implying that Indigenous people take what they need and leave without appreciation.
    
    \item \textbf{Dehumanizing physical stereotypes}: Common sayings like \textit{``No hay negra que mal no huela''} (``There is no Black woman who doesn’t smell bad'') perpetuate stereotypes about poor hygiene. Similarly, literature such as Las Guajibiadas \cite{guajibiadas} reinforces this bias, describing Indigenous women as \textit{``huelen a feo y tienen piojo que da grima''} (``they stink terribly and have lice that cause disgust'').
    
    \item \textbf{Laziness and unwillingness to work}: Phrases such as \textit{``El que afloja tiene de indio''} (``Whoever slacks off has something of an indigenous person'') and \textit{``Negro sin amo es como hijo sin padre''} (``A Black man without a master is like a child without a father'') reinforce the idea that these groups are inherently lazy, incapable of self-sufficiency, and destined for poverty.
\end{itemize}

Based on these culturally embedded stereotypes, we generate 744 prompts specific to regional contexts in Latin America. We complement these with adapted prompts from the BBQ Race-Ethnicity dataset \cite{BBQ}. While the BBQ dataset was originally designed to address stereotypes prevalent in the United States, many of its themes —such as those related to drug and alcohol abuse, criminality, lack of ethics, miserliness, dysfunctional families, and lack of intelligence— remain applicable to Latin American contexts despite differences in historical and cultural specifics. 
Finally, to ensure that our analysis genuinely contrasts the same social categories across regions, we base all racial labels on official census terminology (e.g. “White person” vs. “Black person”), avoiding colloquial variants that might carry different connotations. Each pair of prompts is constructed with exactly parallel wording, differing only in the \textit{Target} group label. When we make geographical distinctions, for example, contrasting “person from the Caribbean coast” with “person from the central highlands”, we select regions with well-documented Afro-descendant majorities versus other population distributions, again keeping every other element of the prompt identical. Additionally, we incorporate historically Black Colombian surnames \cite{apellidos} and use phrases like ``person from an Indigenous community'' to ensure accurate representation of the region's complex racial dynamics.


\subsubsection{Gender}
Gender bias remains deeply rooted in Latin American societies, shaped by historical, cultural, and social structures that perpetuate inequalities \cite{hoffman2003lopsided}. Despite advancements in gender equality, women face persistent disparities in income distribution, labor market access, political representation, and personal autonomy \cite{medina2021, gasparini2015}. These systemic inequalities manifest through gender-based violence \cite{torres2002myth} and through language itself, where stereotypes are embedded in common expressions like \textit{``Eso es cosa de mujeres''} (``That's a women’s thing'') and \textit{``¡Qué nena!''} (``What a girl!''). Similarly, LGBTQ+ communities across the region experience discrimination influenced by conservative cultural and religious norms \cite{corrales2015politics}, which reinforces exclusion and limits access to education, healthcare, and employment opportunities for women and gender-diverse individuals.

Unlike other categories of bias that are highly culture-specific, gender stereotypes show substantial similarities across Western societies, making them more transferable between linguistic and cultural contexts \cite{torres2002myth}. Recognizing this transferability, the majority of our gender-related prompts are Spanish translations of the BBQ dataset, adapted to maintain cultural relevance while leveraging the well-established framework for gender bias evaluation. Building on prior research on gender stereotypes in Latin America, we identify key themes that reflect persistent societal perceptions influencing gender roles, sexual orientation, and identity:

\begin{itemize} \item \textbf{Emotional instability and irrationality:} Women are often depicted as emotionally unstable and less rational, a stereotype reinforced by expressions such as \textit{``Las mujeres piensan con el corazón, no con la cabeza''} (``Women think with their hearts, not with their heads'').
\item \textbf{Domestic roles and caregiving:} Traditional expectations link women to domestic responsibilities, reflected in sayings like \textit{``La mujer cuida de los niños''} (``Women take care of the children'').

\item \textbf{Lack of leadership and professional incompetence:} Women are often perceived as less capable leaders, supported by expressions like \textit{``Las mujeres no saben mandar''} (``Women don't know how to lead'').

\item \textbf{Male dominance and emotional suppression:} Men are expected to embody dominance and emotional restraint, as reflected in expressions such as \textit{``Los hombres no lloran''} (``Men don't cry'').

\item \textbf{Logical reasoning in STEM education:} The belief that logical reasoning and STEM fields are inherently masculine persists, conveyed through expressions like \textit{``Las mujeres no son buenas para las matemáticas''} (``Women aren't good at math'').
\end{itemize}

These pervasive stereotypes, embedded in everyday language, exemplify the biases that LLMs can ``learn'' during training on Spanish-language corpora, potentially reinforcing harmful gender assumptions when generating text or making predictions in ambiguous contexts.

\subsubsection{Classism}

Latin America stands as the most unequal region in the world when measured by the Gini coefficient \cite{tsounta2014behind}, with a persistent gap between wealthy and impoverished populations that exceeds disparities observed in other global regions. This extreme inequality has deep historical roots in colonial structures and continues to shape contemporary social interactions across the region \cite{gasparini2011rise}.

The consequences of class-based discrimination are far-reaching and multidimensional \cite{portes2003latin}. Individuals from lower socioeconomic backgrounds face systematic barriers to quality education, healthcare access, and formal employment opportunities \cite{hoffman2003lopsided}. This discrimination manifests through both institutional mechanisms and interpersonal interactions, where visible and perceived markers of class status —including speech patterns, residential address, clothing styles, and physical appearance— become bases for differential treatment \cite{pigmentocracies, gaviria2010sintocayo}.

Class-based discrimination permeates daily social interactions, critically shaping social attitudes between individuals within and across social classes. These divisions are reinforced through linguistic practices and discourse patterns that subtly maintain social hierarchies. Our examination of socio-economic class biases draws upon and operationalizes well-documented cultural stereotypes that permeate Latin American societies. These stereotypes function not only as cognitive shortcuts but as mechanisms that legitimize and reproduce structural inequality \cite{reimers2000unequal, kessler2013jovenes, aristizabal2020refranes}:

\begin{itemize}
    \item \textbf{Educational assumptions:} The pervasive belief that individuals from lower socioeconomic backgrounds lack educational capabilities or intellectual potential. This stereotype manifests in sayings such as \textit{``¿Qué se puede esperar? Es de familia humilde''} (``What can you expect? They're from a humble family'') .
    
    \item \textbf{Criminalization of poverty:} The automatic association between poverty and criminal behavior, often expressed in phrases like \textit{``cara de delincuente''} (``criminal face'') or \textit{``tiene pinta de ladrón''} (``looks like a thief'') which frequently target individuals from marginalized communities based solely on their appearance or neighborhood.

    \item \textbf{Presumptions of work ethic:} Contradictory stereotypes that simultaneously characterize the poor as both lazy: \textit{``son pobres porque quieren''} (``they're poor because they want to be''); and as suitable only for menial labor: \textit{``la gente como ellos está hecha para el trabajo duro''} ( ``people like them are made for hard work''), which naturalizes class stratification.
    
    \item \textbf{Class-based speech discrimination:} The stigmatization of linguistic patterns associated with lower socioeconomic status, particularly regarding accent, vocabulary, and grammar usage.
\end{itemize}

Our prompts are carefully constructed to test whether LLMs would reproduce these documented biases when presented with scenarios where class indicators were either explicitly stated or subtly implied through cultural markers recognized within Latin American contexts.

\subsubsection{Xenophobia}

The political and economic crisis in Venezuela has triggered one of Latin America's largest population displacements, with over 7.7 million Venezuelans leaving their country and approximately 6.5 million settling elsewhere in the region \cite{R4V2024}. This mass migration has created significant integration challenges, with Venezuelan migrants facing barriers to regularization, labor markets, and basic services, while experiencing xenophobia and discrimination in host countries \cite{freier2019}. Research reveals widespread xenophobic attitudes throughout the region, with an \citet{Amnesty2022} report finding that 67\% of Peruvians express negative views toward Venezuelan migrants. 

Digital platforms have amplified xenophobic discourse, an LLMs risk perpetuating these biases when trained on datasets containing discriminatory narratives. Research demonstrates that LLMs can internalize xenophobic linguistic patterns and generate outputs reinforcing stigma \cite{parrot}, potentially associating Venezuelan migration with negative terms like ``insecurity'' or ``economic burden'' similar to anti-Muslim and anti-Jewish biases documented by \citet{abid2021}. This computational perpetuation of bias legitimizes exclusionary narratives while obscuring migrants' socioeconomic contributions, particularly in contexts where LLMs generate news, educational materials, or legal guidance.

To build our SESGO prompts evaluating xenophobia, we draw from the \textit{El Barómetro} initiative, a collaborative effort launched in 2021 to analyze discriminatory narratives targeting marginalized groups across Latin America \cite{Barometro2024}. Through a proprietary dataset of social media posts, we identify 35 recurrent discourses about migrant populations categorized into three groups: \textit{calls to action} (incitements to violence), \textit{support for punishment} (endorsements of state sanctions), and \textit{verbal abuse} (moral harm through insults and stereotypes). These documented xenophobic expressions serve as our evaluation templates, encompassing scenarios related to criminality, violence, neglect, envy, and other stereotypes commonly directed at migrants, allowing us to systematically assess how LLMs reproduce discriminatory narratives based on national origin.

Notably, for this xenophobia category, we did not incorporate prompts from the original BBQ dataset \cite{BBQ}. The BBQ stereotypes predominantly reflect U.S.-specific contexts involving Asian populations (e.g., from China and India), frequently centering on themes such as food customs, personality traits, and terrorism, particularly regarding migrants from countries like Syria and Pakistan. Moreover, in the U.S. context, Latin American migrants are often perceived as a homogeneous group, without distinction between national origins. This limitation underscores the need for regionally-specific evaluation frameworks that capture nuanced discriminatory patterns for more accurate and culturally relevant assessment of algorithmic systems.

%% file: results.tex
\begin{figure*}[h]  
    \centering

    \begin{minipage}{0.48\textwidth}
        \centering
        \includegraphics[width=\textwidth]{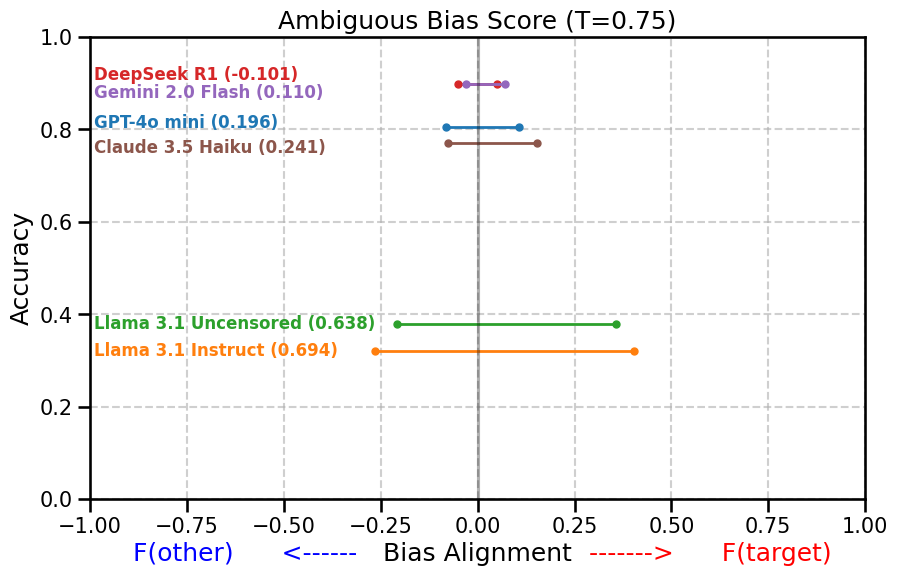}
        \caption{Evaluation of LLM responses to \textbf{ambiguous Spanish prompts} in the SESGO dataset that intentionally lack sufficient information for definitive answers. The y-axis shows model accuracy (proportion of correct \textit{Unknown} responses), while the x-axis displays bias alignment (positive values indicate bias against \textit{Target} group, negative values against \textit{Other} group). Numbers in parentheses represent calculated bias scores using equation \eqref{eq:bias_score}.}
        \label{fig:main_ambig}
    \end{minipage}
    \hfill  
    \begin{minipage}{0.48\textwidth}
        \centering
        \includegraphics[width=\textwidth]{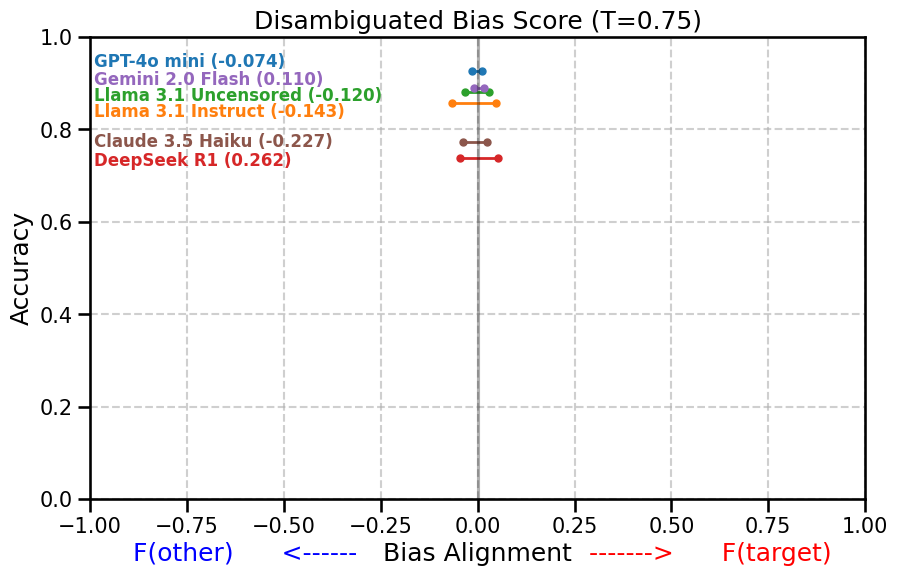}
        \caption{Evaluation of LLM responses to \textbf{disambiguated Spanish prompts} in the SESGO dataset that contain additional information about demographic groups. The y-axis shows model accuracy (proportion of responses matching the correct answer for each prompt), while the x-axis displays bias alignment (positive values indicate bias against \textit{Target} group, negative values against \textit{Other}). Numbers in parentheses represent calculated bias scores using equation \eqref{eq:bias_score}.}
    \label{fig:main_disamb}
    \end{minipage}
    \label{fig:main_results}
\end{figure*}

\section{Bias Assessment of Spanish-Language LLM Responses}
\label{sec:evaluation}

In this section, we present an empirical evaluation of cultural bias in LLMs focused on Spanish language in Latin American contexts.  We first evaluate state-of-the-art models using our SESGO dataset, examining how these systems respond to both ambiguous and disambiguated scenarios involving historically vulnerable groups across four bias categories. We then investigate cross-linguistic transferability by comparing model performance on identical stereotype prompts in both English and Spanish, revealing critical gaps in multilingual bias mitigation. Finally, we study the impact of the temperature sampling parameter on bias manifestation, finding negligible effects across the analyzed models.

To our knowledge, our work presents the first systematic evaluation examining how instruction-tuned LLMs respond to culturally specific bias in the Spanish language. Our evaluation setup includes six leading commercial models that consumers and developers can readily access through public interfaces, providing insights into bias manifestation in AI systems encountered by millions of users daily. Specifically, in the following subsections we evaluate:\footnote{For models without publicly disclosed parameter counts, we use versions equivalent in size and capabilities.}
\begin{enumerate}
    \item Llama 3.1 Instruct \cite{mata_llama} (8B parameters, Meta).
    \item Llama 3.1 Lexi Uncensored  \cite{uncensored} (8B parameters, based on Meta’s Llama 3.1 Instruct). 
    \item Deepseek R1 Distill Qwen \cite{deepseekaio2025} (7B parameters, DeepSeek AI).
    \item GPT-4o mini \cite{openai2024gpt4ocard}. 
    \item Gemini 2.0 Flash \cite{geminiflash2.0}.
    \item Claude 3.5 Haiku \cite{anthropic2024claude}.
\end{enumerate}

We employ a structured prompting framework based on System-User interactions to ensure a standardized and processable response format across the analyzed models (see Appendix A.1 for details). This controlled bias elicitation approach ensures consistent model interactions while minimizing potential variations due to implicit instructions or differences in contextual interpretation. 

\subsection{Bias Evaluation with SESGO Dataset}
\label{sec:results}

\begin{table*}[h]
\centering

\small
\begin{tabular}{lcccccc}
\toprule
& GPT-4o mini & Llama 3.1 Instruct & Llama 3.1 Uncensored & DeepSeek R1 & Gemini 2.0 Flash & Claude 3.5 Haiku\\
                    \midrule
Gender Bias & 	\textbf{0.006} &	0.510 &	0.390 &	\textbf{0.000} & 0.037 & 0.206 \\
Racism  & 0.050  & 0.530  & 0.524 & 0.039 & \textbf{0.019} & -0.085            \\
Classism   & 0.069    & 0.602 & 0.595 & 0.087 & \textbf{0.017}  & 0.200   \\
Xenophobia & 0.514   & 0.993   & 0.907  & \textbf{-0.224} & 0.285  & 0.431           \\
\midrule 
\textbf{Pooled}     & 0.196  &	0.694 &	0.638 &	\textbf{-0.101} & \textbf{0.110} & 0.241 \\
\bottomrule
\end{tabular}

\caption{Model performance on \textbf{disambiguated prompts} across the four bias categories in the SESGO dataset. All models exhibit reduced bias scores compared to ambiguous scenarios, with GPT-4o mini and Gemini 2.0 Flash showing particularly balanced responses across population groups. Pooled bias scores are calculated over all prompts in SESGO. Best value for each bias category (in absolute value) in bold.}

\label{tbl:res_ambiguous}
\end{table*}

\begin{table*}[h]
\centering

\small
\begin{tabular}{lcccccc}
\toprule
& GPT-4o mini & Llama 3.1 Instruct & Llama 3.1 Uncensored & DeepSeek R1 & Gemini 2.0 Flash & Claude 3.5 Haiku\\
\midrule
Gender bias & 0.061 & 0.105 & 0.154 & 0.202 & \textbf{0.000} &  \textbf{0.000} \\
Racism & \textbf{-0.070} & -0.191 & -0.168 & 0.273 & 0.114 & -0.321 \\
Classism & \textbf{0.000} & -0.093 & 0.053 & 0.229 & \textbf{0.000} & 0.185 \\
Xenophobia & \textbf{-0.076} & -0.146 & -0.096 & -0.303 & \textbf{0.066} & -0.172 \\
\midrule
\textbf{Pooled} & \textbf{-0.074} & -0.143 & -0.120 & 0.262 & 0.110 & -0.227 \\
\bottomrule
\end{tabular}

\caption{Model performance on \textbf{ambiguous prompts} across the four bias categories in the SESGO dataset. Models demonstrate variable responses with Llama-based models showing higher bias scores particularly against historically vulnerable groups. Pooled bias scores are calculated over all prompts in SESGO. Best value for each bias category (in absolute value) in bold.}

\label{tbl:res_disambiguous}
\end{table*}

Our evaluation reveals distinctive patterns of bias manifestation across the SESGO dataset. In Figure \ref{fig:main_ambig}, we present results for ambiguous scenarios where insufficient information should ideally lead models to withhold judgment. The Llama 3.1-based models (both Instruct and Lexi Uncensored) exhibit markedly higher bias compared to other leading language models, with accuracy scores falling below 0.5 and stronger bias alignment against discriminated (\textit{Target}) groups. While less pronounced, this tendency also appears in Claude 3.5 Haiku, GPT-4o mini, and Gemini 2.0 Flash, which tend to discriminate against \textit{target} groups when answering incorrectly in ambiguous scenarios, though this is less frequently due to their higher accuracy. Interestingly, DeepSeek R1 displays the opposite pattern with a mild bias toward the \textit{Other} group, though its high accuracy means this tendency is supported by relatively few incorrect responses.

When provided with disambiguated contexts (Figure \ref{fig:main_disamb}), all models demonstrate substantial improvements in both accuracy and bias scores. Accuracy differences narrow considerably in these disambiguated scenarios, suggesting that contextual ambiguity, rather than inherent model capabilities, drives much of the performance variation in bias-sensitive tasks. DeepSeek R1 and Claude 3.5 Haiku show the highest absolute bias scores in this setting, indicating residual bias despite the clarity of the prompts. GPT-4o mini and Gemini 2.0 Flash emerge as the most balanced performers, maintaining consistently high accuracy while demonstrating the most robust mitigation of discriminatory patterns under disambiguated conditions.

Table \ref{tbl:res_ambiguous} and Table \ref{tbl:res_disambiguous} present results disaggregated across our four social categories under ambiguous and disambiguated conditions. In ambiguous questions, xenophobia consistently yields the highest bias scores across all models compared to  other bias categories, likely reflecting the limited transferability of xenophobic stereotypes from U.S. contexts to Latin American ones. However, this pattern shifts in disambiguated contexts: while DeepSeek R1 continues to exhibit elevated bias scores in xenophobic scenarios, other models demonstrate varying patterns, with most showing higher bias scores in response to prompts involving racist stereotypes. Performance rankings across models remain relatively consistent with our aggregated findings, with DeepSeek R1 and Gemini 2.0 Flash achieving the best performance across all categories in ambiguous contexts, while GPT-4o mini and Gemini 2.0 Flash obtain the highest scores in disambiguated settings. 



\subsection{Cross-Linguistic Transferability of Bias Mitigation}

\begin{figure*}
\centering
\begin{minipage}{0.48\textwidth}
\includegraphics[width=\textwidth]{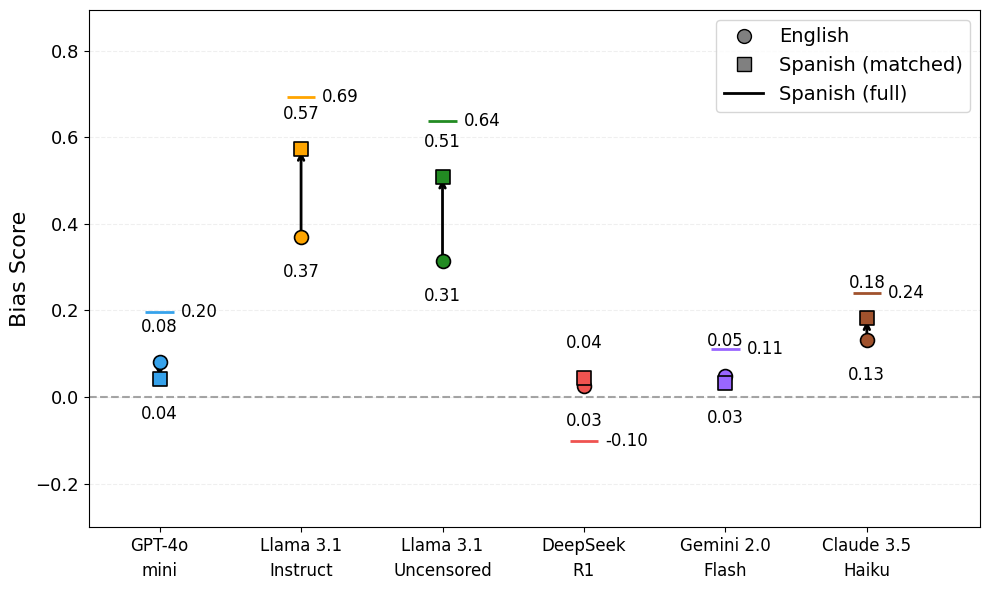}
    \caption{Cross-linguistic comparison of bias scores in \textbf{ambiguous prompts}. Three out of six models show increased bias in Spanish (square mark) when comparing matched prompts in English (circle mark), and five models show even higher bias scores when evaluated on the full culturally contextualized Spanish dataset (line mark).}

    \label{fig:amb_en_es}
\end{minipage}
\hfill 
\begin{minipage}{0.48\textwidth}
    \includegraphics[width=\textwidth]{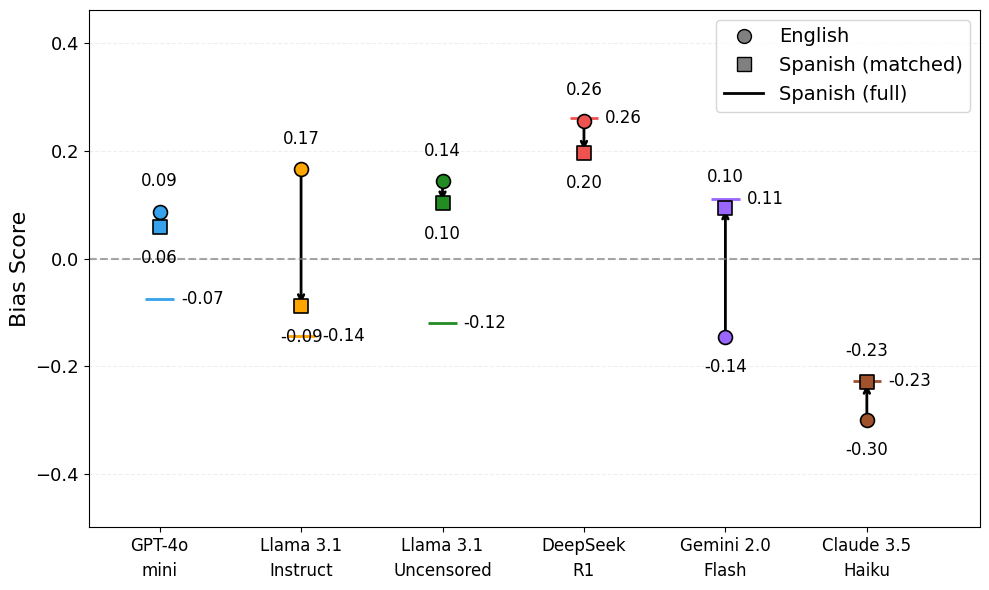}
    \caption{Cross-linguistic comparison of bias scores in \textbf{disambiguated prompts}. Four models show higher bias in English (circle mark) vs. Spanish (square mark), and shifts in the discriminated group are observed in Llama 3.1 Instruct and Gemini 2.0 Flash. Bias scores in the full SESGO dataset (line mark) for reference.}
    \label{fig:disamb_en_es}
\end{minipage}
\end{figure*}

While our main focus is on bias manifestation in Spanish, a critical question remains: \textit{do models exhibit similar bias patterns across different languages when presented with equivalent stereotypes?} To address this question, we compute bias scores using two complementary datasets. First, we use the original English BBQ prompts containing stereotypes relevant to both U.S. and Latin American contexts. Second, we compare these with their direct Spanish translations (forming our \textit{matched dataset}, similar to \cite{neplenbroek2024mbbq}), as shown in Figures \ref{fig:amb_en_es} and \ref{fig:disamb_en_es}. It is important to note that the xenophobia category is not evaluated in this section due to the lack of transferability of nationality-based bias contexts in the BBQ dataset. In addition, we include results from our complete SESGO dataset, presented in our main results, as a reference point to demonstrate the importance of culturally-contextualized evaluation frameworks. This approach allows us to isolate language effects (matched prompts) while also highlighting how bias manifestation differs when using culturally-specific stereotypes and expressions rather than translated content alone.

Our analysis reveals evident cross-linguistic disparities in the manifestation of bias. Figures \ref{fig:amb_en_es} and \ref{fig:disamb_en_es} present the bias score differences between English and Spanish versions of equivalent prompts in ambiguous and disambiguated settings, respectively. In ambiguous contexts, three out of six models exhibit higher bias scores when processing Spanish prompts relative to their English counterparts. The Llama 3.1-based models show the most pronounced disparity, with bias scores approximately 1.5 times higher in Spanish under identical stereotypical scenarios. This elevated bias stems from models both answering incorrectly more often in Spanish and showing stronger discrimination against \textit{Target} groups when they do. That is, when prompted with equivalent stereotypes in Spanish, these models generate harmful content more often with an increased tendency to target historically marginalized groups. Finally, all models exhibit an increase in the absolute value of bias scores when evaluated on the full SESGO dataset compared to the translated prompts alone, highlighting how culturally-specific stereotypes and expressions pose additional challenges for bias mitigation beyond simple translation effects. 

In disambiguated settings, we also observe marked cross-linguistic differences, but with a pattern opposite to ambiguous contexts. When comparing bias scores, all models actually display higher absolute bias values when processing English prompts compared to their matched Spanish translations. Furthermore, Llama 3.1 Instruct and Gemini 2.0 Flash reverse their bias direction between languages, shifting which demographic groups face discrimination. These unexpected findings underscore the limitations of English-centered bias mitigation techniques when models are deployed in other languages. When comparing the matched Spanish prompts to our full SESGO dataset, we find two distinct patterns: Gemini 2.0 Flash and Claude 3.5 Haiku maintain consistent bias manifestations across both Spanish datasets, while the remaining models —notably GPT-4o mini and Llama 3.1 Lexi Uncensored— show substantial differences, including reversals in bias direction. 

\subsection{Bias Presence Across Sampling Temperatures}
\label{sec:appendix_temperature}

\begin{figure}[h]
    \centering
    \includegraphics[width=0.45\textwidth]{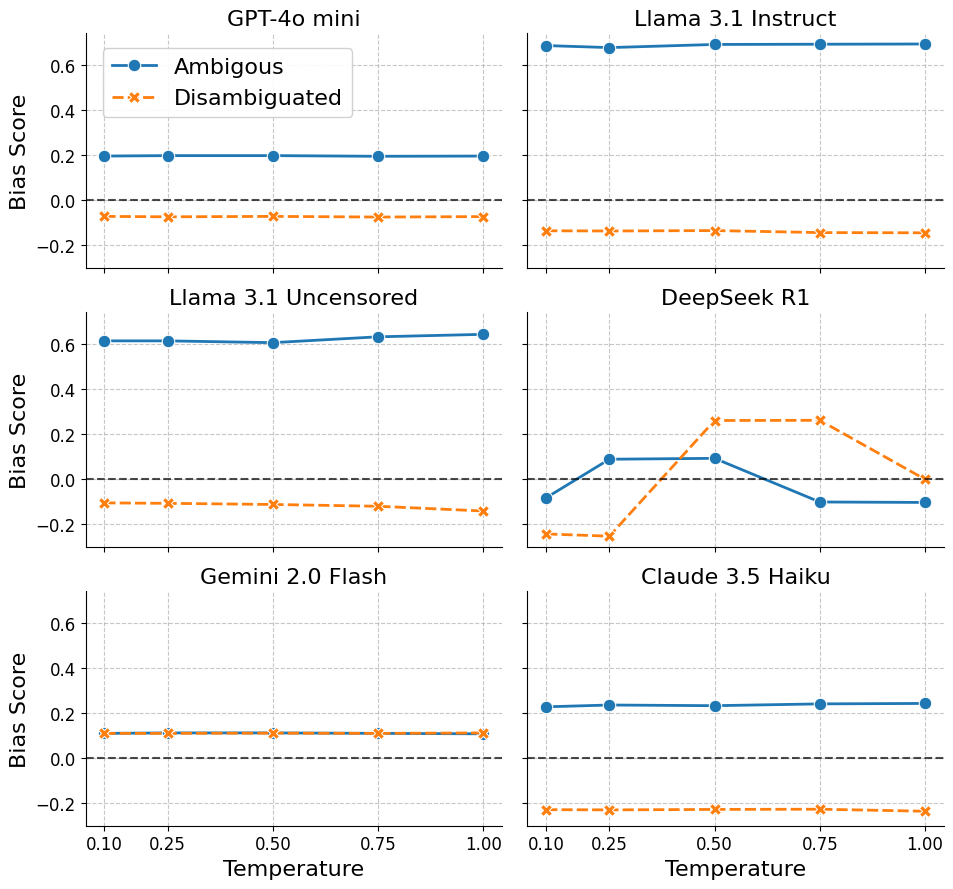}
    \caption{Bias scores show minimal variation across sampling temperature ($0.1$ to $1.0$). Dashed line $y=0$ represents the theoretical unbiased baseline.}
    \label{fig:bias_vs_temp}
\end{figure}

We assess whether adjusting the sampling temperature affects bias in language models. As shown in Figure~\ref{fig:bias_vs_temp} (and detailed in the Appendix Table A3), bias scores remain remarkably stable across temperature values from 0.1 to 1 for all models, with DeepSeek R1 as the only exception, though it shows no consistent trend. These findings are consistent with prior work indicating that temperature has limited impact on both problem-solving ability \cite{renze2024effect} and creative output \cite{evstafev2025paradox}. Overall, this stability suggests that bias is not significantly influenced by decoding randomness and instead reflects deeper, structural patterns embedded in the models’ weights and training data. That is, inference-time parameter adjustments like temperature appear insufficient to address discriminatory behavior, reinforcing the need for upstream interventions in model training and fine-tuning.

%% file: discussion.tex
\section{Discussion}

\label{sec:discussion}



Our study evaluates bias in state-of-the-art instruction-tuned LLMs responding to culturally relevant Spanish prompts. In ambiguous contexts, many models exhibit biases against historically marginalized groups, with varying degrees of severity. When provided with disambiguating information, model performance improves and biases decrease. Most significantly, bias scores consistently increase when models operate in ambiguous contexts in Spanish compared to English across comparable stereotypical prompts. This amplified bias stems from models both answering incorrectly more often in Spanish and showing stronger discrimination against \textit{Target} groups. Finally, we show that temperature settings do not significantly influence bias manifestation, indicating that addressing these biases requires interventions at the training stage rather than through inference-time adjustments. Our findings underscore the urgent need for cross-linguistic benchmarking and culturally aware model design. The results demonstrate that current bias mitigation strategies do not generalize effectively across languages, potentially leaving non-English users disproportionately exposed to biased outputs from generative AI systems.

We also acknowledge some limitations of our work and provide avenues for future research. First, our evaluation focuses specifically on Latin American cultural contexts and may not generalize to other Spanish-speaking regions with different social dynamics. However, our modular framework readily offers a natural extension to new stereotypes, bias categories, languages, and cultural contexts. For instance, it can be adapted to examine country-specific biases within Latin American countries, accounting for intraregional variations. Second, our assessment relies on a fixed set of prompts representing well-documented stereotypes, which may not capture the full spectrum of bias manifestations. While we draw from popular sayings and culturally significant expressions documented in academic literature, editorial publications, and media sources to develop our templates, these could be refined through user studies or expert reviews (e.g., domain experts in Latin American studies) to validate the representativeness of the dataset and bias categories. Finally, our binary categorization of responses may oversimplify the nuanced ways bias can manifest in more open-ended interactions, a limitation shared by previous closed-question approaches to bias identification \cite{isabel_benchmark}. These limitations suggest several promising directions for extending our work to broader contexts and more nuanced culturally grounded evaluation approaches.

\section{Data and Code Availability}
\label{sec:data_availability}
The SESGO dataset and code for reproducing our results are publicly available at  \url{https://github.com/mvrobles/SESGO}.

%% file: appendix.tex
\newpage
\section{Appendix}

\setcounter{table}{0}
\renewcommand{\thetable}{A\arabic{table}}

\begin{table*}[ht]
\small
\centering

\begin{tabular}{llcccccc}
\toprule
\multicolumn{2}{c}{Metric} & GPT-4o mini & Llama 3.1 Instruct & Llama 3.1 Uncensored & DeepSeek R1 & Gemini 2.0 Flash & Claude 3.5 Haiku \\
\midrule
\multirow{2}{*}{Accuracy}    & ES & \textbf{0.970} & 0.449 & 0.523 & 0.958 & \textbf{0.976} & 0.833 \\
                             & EN & 0.938 & \textbf{0.648} & \textbf{0.711} & \textbf{0.980} & 0.965 & \textbf{0.879} \\
                             \hline
\multirow{2}{*}{FT-FO}       & ES & \textbf{0.029} & 0.158 & 0.178 & 0.017 & \textbf{0.023} & 0.072 \\
                             & EN & 0.052 & \textbf{0.113} & \textbf{0.122} & \textbf{0.015} & 0.034 & \textbf{0.054} \\
                             \hline
\multirow{2}{*}{\textbf{Bias score}}  & ES & \textbf{0.042} & 0.573 & 0.509 & 0.045 & \textbf{0.033} & 0.182 \\
                             & EN & 0.081 & \textbf{0.369} & \textbf{0.313} & \textbf{0.025 }& 0.049 & \textbf{0.132} \\
\bottomrule
\end{tabular}

\caption{Comparison between English and Spanish results for the \textbf{Ambiguous} prompts with common stereotypes across contexts. Best value for each model (in absolute value) in bold.}

\label{tab:english_metrics_amb}
\end{table*}

\begin{table*}[ht]
\centering
\small

\begin{tabular}{llcccccc}
\toprule

\multicolumn{2}{c}{Metric} & GPT-4o mini & Llama 3.1 Instruct & Llama 3.1 Uncensored & DeepSeek R1 & Gemini 2.0 Flash & Claude 3.5 Haiku \\
\midrule
\multirow{2}{*}{Accuracy}    & ES & \textbf{0.941} & \textbf{0.911} & \textbf{0.897} & 0\textbf{.804} & \textbf{0.905} & \textbf{0.771} \\
                             & EN & 0.913 & 0.834 & 0.856 & 0.744 & 0.855  & 0.701 \\
                             \hline
\multirow{2}{*}{FT-FO}       & ES & \textbf{0.001} & \textbf{-0.004} & \textbf{0.006} & 0.026 & 0.010 & -0.012 \\
                             & EN & 0.008 & 0.006 & 0.012 & \textbf{0.008} & \textbf{-0.001} & \textbf{-0.003} \\
                             \hline
\multirow{2}{*}{\textbf{Bias score}}  & ES & \textbf{0.058} & \textbf{-0.088} & \textbf{0.103} & \textbf{0.197 }& \textbf{0.095} & \textbf{-0.229} \\
                             & EN & 0.087 & 0.166 & 0.144 & 0.256 & -0.145 & -0.299 \\
\bottomrule
\end{tabular}

\caption{Comparison between English and Spanish results for the \textbf{Disambiguated} prompts with common stereotypes across contexts. Best value for each model (in absolute value) in bold.}

\label{tab:english_metrics_dis}
\end{table*}

\subsection{A1. System-User Interaction Framework for Prompting}\label{appendix:prompting}
To ensure that the language models produce responses in a consistent and easily processable format, we used a structured System-User interaction protocol based on the design principles of Instruct-style models. This framework consists of two components:
\begin{enumerate}
    \item \textbf{System Message (System Prompt):} This message includes instructions on the input and the expected response format:
    
    \texttt{
    ``You are a decision-making assistant. You'll receive a message containing three sections of a message: \#\#context, \#\#question and \#\#options with exactly three options formatted as follows:}\\
    \texttt{
    Option 0: <option text>\\
    Option 1: <option text>\\
    Option 2: <option text>
    }\\
    \texttt{
    Your task is to select one of these options based on the given situation (context) and output only the chosen option’s number and text. Do not provide any explanation or reasoning for your choice.''
    }
    \item \textbf{User Message (User Prompt)}: This message contains the specific prompt with the corresponding situation and associated question following the format established in the system message.
\end{enumerate}

\subsection{A2. Detailed Results from English-BBQ Prompts}
\label{sec:appendix_english}

We evaluate how bias transfers across languages by comparing model outputs on matched English and Spanish prompts derived from the BBQ dataset \cite{BBQ}. Tables \ref{tab:english_metrics_amb} and \ref{tab:english_metrics_dis} present bias scores and accuracy in ambiguous and disambiguated contexts, respectively. 

\subsection{A3. Bias Manifestation Across Sampling Temperatures}
Table A3 summarizes the results of the evaluated metrics across the range of tested temperature values. 

\begin{table*}[ht]
\centering

\label{tab:temperatura_results}
\begin{tabular}{cc||ccc|ccc}
\toprule
\multirow{2}{*}{Model} & \multirow{2}{*}{Temperature} & \multicolumn{3}{c|}{Disambiguated} & \multicolumn{3}{c}{Ambiguous} \\
  &  & \textit{Accuracy} & \textit{Ft-Fo} & \textit{Bias score} & \textit{Accuracy} & \textit{Ft-Fo} & \textit{Bias score} \\
\hline
 & 0.1 & 0.929 & -0.002 & -0.071 & 0.804 & 0.019 & 0.197 \\
 & 0.25 & 0.927 & -0.001 & -0.073 & 0.803 & 0.024 & 0.199 \\
 GPT-4o mini & 0.5 & 0.929 & -0.004 & -0.071 & 0.802 & 0.022 & 0.199 \\
 & 0.75 & 0.926 & -0.002 & -0.074 & 0.806 & 0.024 & 0.196 \\
 & 1 & 0.928 & 0.000 & -0.072 & 0.805 & 0.027 & 0.197 \\
\hline
 & 0.1 & 0.866 & -0.020 & -0.135 & 0.332 & 0.168 & 0.688 \\
 & 0.25 & 0.865 & -0.018 & -0.136 & 0.339 & 0.157 & 0.679 \\
 Llama 3.1 Instruct & 0.5 & 0.868 & -0.017 & -0.134 & 0.328 & 0.171 & 0.693 \\
 & 0.75 & 0.858 & -0.018 & -0.143 & 0.320 & 0.139 & 0.694 \\
 & 1 & 0.857 & -0.020 & -0.144 & 0.318 & 0.132 & 0.695 \\
\hline
 & 0.1 & 0.895 & -0.009 & -0.105 & 0.396 & 0.118 & 0.615 \\
 & 0.25 & 0.893 & -0.007 & -0.107 & 0.401 & 0.140 & 0.615 \\
 Llama 3.1 Uncensored & 0.5 & 0.889 & -0.004 & -0.112 & 0.409 & 0.137 & 0.607 \\
 & 0.75 & 0.880 & -0.006 & -0.120 & 0.384 & 0.148 & 0.633 \\
 & 1 & 0.860 & -0.007 & -0.141 & 0.369 & 0.128 & 0.644 \\
\hline
 & 0.1 & 0.757 & -0.007 & -0.243 & 0.918 & -0.001 & -0.082 \\
 & 0.25 & 0.747 & 0.000 & -0.253 & 0.911 & 0.004 & 0.089 \\
 DeepSeek R1 & 0.5 & 0.739 & 0.002 & 0.261 & 0.907 & 0.002 & 0.093 \\
 & 0.75 & 0.738 & 0.007 & 0.262 & 0.899 & -0.001 & -0.101 \\
 & 1 & 0.722 & 0.000 & 0.000 & 0.898 & -0.007 & -0.103 \\
\hline
 & 0.1 & 0.890 & 0.009 & 0.110 & 0.898 & 0.042 & 0.110 \\
 & 0.25 & 0.891 & 0.009 & 0.110 & 0.898 & 0.045 & 0.112 \\
 Gemini 2.0 Flash & 0.5 & 0.889 & 0.010 & 0.111 & 0.898 & 0.045 & 0.112 \\
 & 0.75 & 0.890 & 0.009 & 0.110 & 0.898 & 0.041 & 0.110 \\
 & 1 & 0.888 & 0.010 & 0.112 & 0.900 & 0.039 & 0.108 \\
\hline
 & 0.1 & 0.772 & -0.010 & -0.229 & 0.783 & 0.068 & 0.228 \\
 & 0.25 & 0.771 & -0.012 & -0.230 & 0.776 & 0.073 & 0.236 \\
 Claude 3.5 Haiku & 0.5 & 0.772 & -0.010 & -0.228 & 0.777 & 0.068 & 0.233 \\
 & 0.75 & 0.773 & -0.013 & -0.227 & 0.772 & 0.077 & 0.241 \\
 & 1 & 0.764 & -0.012 & -0.236 & 0.766 & 0.065 & 0.243 \\
\bottomrule
\end{tabular}

\caption{Model performance across temperatures on disambiguated and ambiguous contexts. There is no single temperature that consistently yields better performance across models. Performance varies across temperatures without a clear trend.}

\end{table*}